\documentclass{iau} 
\usepackage{graphicx}

\title[IAU-S334~~Closing remarks and Outlook] 
{Closing remarks and Outlook}

\author[F. Combes]   
{Francoise Combes$^1$}

\affiliation{$^1$Observatoire de Paris, LERMA, College de France, CNRS, PSL, Sorbonne Univ.
UPMC, F-75014, Paris, France \\ email: {\tt francoise.combes@obspm.fr}}

\pubyear{2017}
\volume{334}  
\jname{Rediscovering our Galaxy}
\editors{C. Chiappini, I. Minchev, E. Starkenburg, M. Valentini., eds.}
\begin{document}

\maketitle

\begin{abstract}
Some highlights are given of the IAU Symposium 334, Rediscovering our Galaxy, held
in Potsdam, in July 2017: from the first stars fossil records found in the halo, the 
carbon-enhanced metal poor CEMP-no, to the cosmological simulations presenting
possible scenarios for the Milky Way formation, passing through the chemo-dynamical models
of the various components, thin and thick disks, box/peanut bulge, halo, etc.
The domain is experiencing (or will be in the near future) huge improvements
with precise and accurate stellar ages, provided by astero-seismology, precise stellar distances 
and kinematics (parallaxes and proper motions from GAIA), and the big data resulting
from large surveys are treated with deep learning algorithms.
\end{abstract}


Due to our location well inside the plane, far from the center, the image we have of our 
Galaxy is quite indirect, highly dependent on models, and inspired by potentially
similar external galaxies. Frequently along this week, artist's views of the face-on Milky
Way were shown (often from R. Hurt), but some have two arms and a bar, some have four arms,
more or less winding out. So many different points of view appeared, that it could remind
us the blind scientists examining an elephant, and finding a fan, wall, tree, snake or rope.

\section{Fossil records and halo stars}

The oldest stars, found in the stellar halo, are extremely metal poor, 
only enriched by the first stars in the Universe, the PopIII stars 
which have already disappeared. They are rare, with peculiar abundances.
Most of them are carbon enriched and are dubbed CEMP stars
  (cf J. Norris review talk).

One of the main impact of rotation in massive stars is to produce nitrogen, carbon and some s-process already in the first phases of the chemical enrichment of the Universe.
As Georges Meynet emphasized in his beautiful talk, rotation is 
particularly important for massive low-metallicity stars, since they lack of opacity.
Without rotation, they would only have little stellar winds or mixing. Rotation
helps the elements formed in the stellar core to come to the surface. 
For solar-metallicity stars on the main sequence, rotation makes
very little difference.
Fast-rotating massive stars, or spinstars have been modeled
successfully and can explain the abundances of CEMP stars (Meynet et al 2006, 2010).

The neutron-capture elements (Sr first peak, Ba second peak) are also peculiar,
they might be formed by s or r process. However, most of the
CEMP-s and r/s stars are binaries, their peculiar abundances can be
explained by mass transfer from an AGB star to a lower mass companion.
Eventually, the CEMP-no stars are the best tools for galactic archeology.

Now we learn that these CEMP stars exist also in dwarfs (previously only in massive galaxies).
The r-process elements, formed in rare explosive events, are particularly
over-abundant in Ret II, and only stochastic processes can account for that
(cf the Sculptor dwarf and talk by A. Chiti).

The enrichment of old stars was detailed in G. Cescutti talk:
models of supernovae scenarios with electron capture (EC) or
Magneto-rotation driven (MRD) process were tested. The 
observed stochasticity is well reproduced.
The abundance ratios between
light (Sr \& Y) over heavy (Eu \& Ba) elements require 
spinstars (Cescutti \& Chiappini 2014).
It might be possible to disentangle the various processes, through
the Ba isotopic abundances.

The stellar halo is also very precious to learn about
all accreting events experienced by the MW. Accreted dwarf companions
can be followed through tidal streams and sub-structures in the halo.
In her talk, A. Helmi demonstrated that simulated models
are compatible with the whole halo assembled from companions.
 There are however some discrepancies between the [$\alpha$/Fe] - [Fe/H] 
abundances of halo stars and those of dwarfs today, so not all the stellar
halo could come from this kind of dwarfs.
There could exist two components in the stellar halo,
i.e. an inner \& outer halo (e.g. Carollo et al., 2016).

 The observation of the tidal streams (with TGAS+RAVE) is essential to 
characterize the dark matter halo. Several studies have not converged
yet on its flattening.
Certainly there is hope in the near future from 
the proper motions, that will be determined with high precision by GAIA.
In her talk N. Kallivayalil updated her
measurement of the LMC proper motions, and the coherence 
and accuracy was impressive.
This was crucial to change our view of the Magellanic Clouds orbits, which 
are now known to be on their first passage around the Milky Way.

\section{Amplification from gravitational lenses}

In his talk, T. Bensby described how useful were lenses
to study high resolution spectra of dwarf stars in the
bulge, with high signal to noise ratio. Without gravitational amplification,
it would not have been possible.
Several episodes of star formation have been identified: 3, 6, 8, and 11 Gyr ago
in the bulge. The spectra show several peaks in [Fe/H] over an otherwise wide 
metallicity distribution. For the low metallicity bulge stars, 
the knee of [alpha/M] appears at higher [Fe/H] than for the local thick disk.
However, the finding of so many young stars in the bulge is in contradiction
 with HST proper motion cleaned CMDs, and with the fact that a counterpart of 
this younger population has not yet been found. Some biases, or contamination
by foreground are possible.

Micro-lenses from OGLE were also used by Wegg et al. (2017) to 
characterize stellar populations towards the bulge. 
With 3000 micro-lensing events, it was possible to constrain the 
masses of the lenses, since they determine the time-scale of lensing.
 Together with the M2M dynamical model, done with O.~Gerhard group, it was
possible to show that the stellar IMF (essentially its low-mass end)
 is constant in the Galaxy, whatever
the age of the populations, or the location (bulge, disk), or the alpha 
over-abundance. 

C. Wegg also discussed the
constraints on the dark matter profile that can be derived. It is pleasant to see
that the micro-lensing experiments, that were unable to find the dark matter in compact
objects they were searching for, now succeed eventually to derive constraints on the dark
matter halo of the Milky Way. The stellar masses in the center constrain the M/L of the 
baryons, and a more accurate dark matter density is obtained. Together with the 
rotational velocity constraints at the solar circle, this leads to the conclusion that 
there must exist a large core in the dark matter distribution.

\section{Abundance surveys of the Milky Way}

Chemical tagging is a precious tool to
disentangle the history and formation of the various components of the Milky Way.
Doubts of its reliability were discussed by R. Smiljanic in his talk, since
already large abundance inhomogeneities can be observed in a given open cluster.
However, the thin and thick disks components can be easily identified in many
surveys (RAVE, APOGEE, GALAH, GES, LAMOST in outer parts, etc.).
Convincing results were shown by J. Holtzman in his talk. In particular
the [$\alpha$/Fe] versus [Fe/H] plots at various radii and heights
above the plane by Hayden et al. (2015)
have been viewed several times during the week, and they are very encouraging.
They show that indeed chemical tagging is possible,
since the alpha versus metallicity diagrams separate
very clearly the thick and thin disk, in agreement with their locations.

D. Minniti presented the survey of variable stars
VVV =VISTA Variables in the Vía Lactea. It is a
public ESO variability survey in the near-infrared of
RRLyrae, Cepheids, and Red Clump stars.
Already 1 billion sources have been studied, over 562 square degrees.

The metallicity maps, decomposed in metal-poor and metal-rich
components, reveal clearly the bulge/thick disk (very compact component, with 
short characteristic radius)
and metal-rich disk, with large characteristic radius. Zoccali et al 2017 (GIBS survey)
show clearly the bimodality between these two metallicities.

Many publications have already appeared on the GAIA-ESO survey;
A. Recio-Blanco described their rich results in her review talk,
embellished by a personal movie.
I will just highlight one of them, the discovery of
low [alpha/M] abundance in some bulge stars, which have similar
chemical properties with the thin disk. This confirms the
complex multi-component character of the bulge  (Recio-Blanco et al 2017).

Annie Robin presented an update of the Besan\c{c}on local model with TGAS+RAVE. This
updated model is based on St\"ackel potentials, and an asymmetric drift in 3D. The
chemical properties of the thick and thin disks are taken into account. Good agreement
is obtained with observations, also for the 
kinematics, although a small V value for the Sun motion is derived
(Robin et al 2017).

\section{Chemo-dynamical models of the Milky Way}

In her talk, Lia Athanassoula convinced us that
disks can form after a major merger (Athanassoula et al 2016).
This is a possible scenario for the Milky Way, if the major merger event
occurred more than 8 Gyr ago.
In their chemodynamical models (Athanassoula et al. 2017) the mono-abundance populations
succeed in reproducing observations.

Chemo-dynamical models of the barred inner galaxy were also
described by Ortwin Gerhard, who showed a point of view corresponding to some convergence
among several dynamical works (e.g. Di Matteo et al. 2014, Debattista et al. 2017):
The classical bulge is very limited ($<$ 5\% in mass), most prominent are the
thick and thin disks, forming a box/peanut bulge. Important are also the effects
of radial migration, mapping the outer disk in the upper box/peanut component.
With these ingredients, it is possible to explain the vertical and radial abundance gradients
observed in the Milky Way. 

In particular a recent model by Fragkoudi et al. 2017 (presented in poster 20)
showed how a thin+thick disk model succeeds in reproducing 
the positive abundance gradient with longitude.
Previously, it was thought that only a negative gradient of a thin disk, then combined 
with radial migration and formation of a box/peanut bulge, was sufficient to create
this positive gradient.
In recent publications, Portail et al (2017a,b) succeeded to reproduce 
the observations, with a made-to-measure model (M2M) and chemical tagging.

The reversal of abundance gradient at high latitude,
attributed by Minchev et al. (2014) to flaring of mono-age disk populations,
 has been confirmed in simulations by
D. Kawata to be due to a flaring of the thin disk at large radii.
In the plane, the abundance gradient is negative.
But at large height z $>$ 2kpc above the plane, it becomes positive.
Since the thick disk disappears relatively quickly with radius, the
abundance corresponds then to the thin disk, which becomes thick,
and makes the continuity with the thick disk at large radii
(Rahimi et al. 2014).

This interpretation is confirmed with an age gradient study, as
presented by M. Martig in her talk. Comparing the radial distribution
of age and [$\alpha$/Fe] shows that alpha/M is indeed a relative indicator of age.
While age decreases with radius and increases with height above the plane,
it converges at large radius and height to values comparable to that of the thin disk
(Martig et al. 2016).

It is possible that several mono-age populations are each flaring, nested in each other.
This will follow the inside-out formation of the thin disk. These multiple
flaring disks, at progressively larger radius, might explain the observed
constant stellar scale-height of external edge-on disks
(Minchev et al. 2015).

More generally, to reconstruct the formation of our Galaxy, it is primordial to
 identify age populations.
A proxy has been to use [Fe/H] and [$\alpha$/Fe], together with other 
abundances, to identify Mono-Abundance populations (MAP), which might be
also mono-age population.
But there are large uncertainties (Minchev et al. 2016).
And even young [$\alpha$/Fe] stars have been found
 (Chiappini et al. 2015, Martig et al. 2015).
In the combined sample CoRoGEE (CoRoT + APOGEE), where ages
are better defined with asteroseismology masses, there is indeed a
reversal of age populations, that radial migration is insufficient
to explain (Anders et al. 2017a,b).

These young [$\alpha$/Fe] star
could have been formed from gas enriched by thick disk stars,
they will constitute an “intermediate thick disk”.
Alternatively they might not be really young, but evolved blue stragglers from the old thick disk population (Jofr\'e et al. 2017).

\section{Ages of stars}

Marcio Catelan reviewed all methods to obtain the ages of stars
(see also Soderblom 2010), in particular nucleo-cosmochronometry, and the
isochrone methods, based on the main sequence
turn off (MSTO)  which correspond to imperfect clocks,
although they are still valuable to find relative ages,
as Haywood et al. (2013) or Bensby et al. (2014) have shown.
In the near-future, there will be large improvements with
GAIA and asteroseismology (CoRoT, Kepler, K2),
see the workshop Chiappini, Montalban \& Steffen (2016).

In the models, there are frequently stars older than the Universe;
about 20 yrs ago, this was solved by the introduction of dark
energy or a cosmological constant. Today, CMB observations
and the Planck collaboration come up with a high precision
age of 13.802 $\pm$ 0.026 Gyr.
Could stellar physics compete with this high precision cosmology?
At present errors on stellar ages are still of the order of Gyr!

But there are bright new perspectives with asteroseismology,
which provide accurate stellar masses, and help to raise degeneracies.
Asteroseismology can be miraculous, when individual frequencies $\nu$i can be determined, 
as Arlette Noels demonstrated in her talk.
When only physical parameters such as L, Teff, Fe/H are known for
stars on the main sequence (MS) or red giants (RG) or red clump (RC),
ages are known with $>$ 80\% uncertainty.
When $\Delta \nu$  and $\nu$max are known from asteroseismology, these
uncertainties become $>$30\% (MS), and 20\% (RG and RC),
but when the individual frequencies $\nu$i are known, these
become 20\% (MS) and 10\%  (RG and RC).
With Plato, it will be possible to follow up oscillations over
large periods of time (e.g. 150 days) and $\nu$i will be determined.
Masses could be determined with 1\% accuracy 
and ages with 3\%, as presented in G. Davies' talk (see e.g. Miglio et al. 2017).

Accurate ages will improve the determination of
star formation histories (SFH). There has been several SFH determinations,
based on age proxies such as [Si/Fe] in the solar neighborhood,
and then computation of chemical models for the whole Galaxy: stellar yield, mass
loss are taken into account, together with the assumption that the inner disk operates in
a quasi closed box model (Haywood et al. 2013, Snaith et al. 2014, 2015). These studies
have concluded to the existence of a quenching episode 8~Gyr ago. SF quenching
around 10 Gyr ago is also the main conclusion of the two-infall model of 
Chiappini et al. (1997), or see also Fuhrmann (2011). This quenching
could be due to a merger, or to the strengthening of a bar (or both).
Bar quenching has been studied by Khoperskov et al. (2017, see poster 42).

E. Bernard presented a different method to obtain SFH,
the global population method, based on 
CMD fitting, from GAIA + TGAS. The depth of the samples
is ensured by the combination of information from Tycho2, Hipparcos, APASS.
Of course these estimations will be much improved with the DR2 release of GAIA in April 2018. A preliminary 
SFH in the solar neighborhood shows very young stars in the recent 2-4 Gyr,
following a rather slow star formation in the old ages.
Towards Baade's window in the bulge, the SFH reveals the contrary: a starburst in the old
time, and a rather quiescent bulge after t=8 Gyr.
This has been built directly from CMD, and radial migration has
not yet been taken into account.

In his talk, F. Anders demonstrated the
existence of 2 thick disks in the Milky Way. 
To deal with the large sample of APOGEE-TGAS,  
the t-SNE method was used for the first time in abundance space,
to take into account 15 chemical abundances.
The method was applied on an alpha/abundance diagram, weeded out 
from blurring migrating stars (those with high excentricity).
A third component clearly stands out, the ''hamr": high alpha metal rich thick disk.
This could mean that the formation of thick and thin disks overlap in time.
Radial migration might also have an effect there.

\section{Gravitational potential of the Milky Way: dark matter}

J. Binney discussed in his talk about the dynamical methods
to model accurately the Milky Way in the GAIA era. Some assumptions
have to be made, of a relaxed body, possessing some
symmetries; all non-axisymmetries will then be dealt with
as perturbations. The method 
decomposes the Milky Way into multiple components
according to age, chemical composition, kinematics, geometry 
(thin and thick disks, bulge). The
modeling is based on action variables, and at each component
corresponds a function f(J) of the integral of motions. 
Detailed agreement is obtained with local data RAVE-TGAS
(e.g. Binney 2017).

H-W. Rix was less optimistic, and also proposed symmetric models, where 
bars, dark matter halo flattening and satellites would be treated as
perturbations. However, we must admit that
the present tidal streams studies have not yet given clues for the shape
of the dark matter halo. Certainly, it is difficult to assume
that these tidal streams are relaxed, and in equilibrium in the potential.
It is not sure that more and better data will improve the situation, 
at least nothing will be easy with GAIA, TGAS-RAVE, when all selection
functions have to be taken into account, etc.

In the mean time, our knowledge of vertical 
perturbations, due to satellites, such as the Magellanic clouds,
or Sgr dwarf, with masses up to 1/10 of that of the Galaxy,
is progressing quickly. There were several talks about these
oscillations of the plane, called ''Galactic Seismology", by
A. Sheffield, A. Quillen, L. Widrow, C. Laporte, and F. Gomez.
 Several stellar ''streams" in the outer disk have been interpreted
as a manifestation of these oscillations: TriAndromeda, 
Monoceros Ring, A13. These might be a common perturbation,
including the warp and subsequent corrugations of the MW disk
(Xu, Newberg et al. 2015).

\section{Cosmological models of  the Milky Way}

In her review talk, C. Scannapieco addressed the $\Lambda$CDM problem of 
angular momentum, which is being solved.
Low angular momentum gas is accreted first, by galaxies with
shallow potential well, easy to eject.
Then later, higher angular momentum gas is accreted, and forms
the disks observed today.
It is also possible to form disks with light bulges, in under-dense environments
where mergers are rare.

She presented CLUES simulations, made to represent the Local Group formation 
and to reproduce that of the MW and M31.
The main difference between the two galaxies is not the environment, but the merger
history is the most important.

About the cusp-core problem, could it be moderated by velocity dispersion?
K. Oman proposed that observations are not tracing the cusp, which is present
in the DM background in dwarf galaxies, because high gas dispersion washes out
the inner peak in the rotation curve. In cosmological simulations of dwarf galaxies
(APOSTLE, Oman et al. 2017), they show dwarfs with high non-circular motions.
The velocity dispersion has a high impact on measured rotation curves,
which then mimic a cored-profile.
Real dwarf galaxies show also strong non-circular motions.

Finally, A. Wetzel unveiled in his talk a very welcome
chemodynamical success!
In the frame of several FIRE simulations, and in particular the  Latte project,
they aim to simulate the MW formation, and 
the production of elements: H, He, C, N, O, Ca, Si, Mg, Na, Fe. It was possible 
in 2 cases to obtain a bimodal [$\alpha$/Fe] -– [Fe/H] diagram,
like the well known feature of our Galaxy, tracing clearly the thin and thick disks.
The gap between the two curves corresponds to a merger at z$\sim$1, suddenly quenching 
star formation and element production, and re-starting a new sequence, with new gas,
in line with the so-called two infall model suggested by Chiappini et al. (1997) 
exactly 20 years ago.

In addition, he showed that the missing satellite problem has received
a beginning of solution, in the destruction of satellites close to 
the Galaxy, when a compact baryonic disk is present:
the presence of the galaxy disk at the center eliminates through 
tidal field many sub-halos within 50 kpc
(Garrison-Kimmel, Wetzel et al. 2017).

\section{The future: space for improvement}

We learned during this week many ways to improve our knowledge,
I just point towards the main ones:

--(i) About old stars and fossil records of MW formation, 
better ages will be obtained in the near future, better abundances, 
larger statistics, and theoretical progress should be made in chemical processes

--(ii) The inner galaxy is not well known, because of distance
and obscuration, together with crowding.  We need better constraints 
on the thick disk and the boxy/peanut bulge, the 
mixture of stellar populations, the nuclear disk or bar.

--(iii) Ages of stars is a domain where progress will be the most
spectacular, due to asteroseismology yielding accurate masses.
Many surveys by GAIA (T, g, Fe/H), CoRoT-GES, RAVE, LSST, and in the long-term Plato
will provide exquisite data.

--(iv) In the domain of the gravitational potential and dynamics of the MW, huge progress
will come from GAIA (distances, 3D velocities), which should bring constraints to the
dark matter halo, and also on galaxy formation. Cosmological simulations will
catch up with more refined baryonic processes.

One special note should be made on the advent of the
Big Data era, and the use of deep learning algorithms.
The present spectroscopic stellar surveys
(high resolution: APOGEE, GALAH, Gaia-ESO,
or low resolution: RAVE, LAMOST) and in the
near future WEAVE, 4MOST, the
photometric surveys (VVV, PanSTARRS, SkyMapper..)
and the GAIA revolution, with
astrometry and proper motions for a billion stars,
all contribute to the deluge of data, requiring new analysis methods.

Deep learning methods have been discussed this week (CANNON, t-SNE)
which is a mark that the domain has become mature. The methods are no
longer based only on the physical processes, the stellar evolution models.
The same occurred in meteorological predictions, and closer to us for instance
for the subtraction of foregrounds in Planck data.

Concerning the very near future, the GAIA DR2 release next April, 
I am sure that you are all in your starting blocks!

Many thanks to Cristina and the organizing committees for
such a wonderful week!


\begin{thebibliography}{}


\bibitem[]{} Anders, F., Chiappini, C., Rodrigues, T. S. et al. 2017a,
\textit{A\&A}, 597, A30
\bibitem[]{} Anders, F., Chiappini, C., Minchev, I. et al. 2017b,
\textit{A\&A}, 600, A70

\bibitem[]{} Athanassoula, E., Rodionov, S. A., Peschken, N., Lambert, J. C.: 2016,
\textit{ApJ}, 821, 90
\bibitem[]{} Athanassoula, E., Rodionov, S. A., Prantzos, N.: 2017
\textit{MNRAS}, 467 L46

\bibitem[]{} Bensby, T., Feltzing, S., Oey, M. S.: 2014, 
\textit{A\&A}, 562, A71

\bibitem[]{} Binney, J., 2017,
in IAU Symposium 330, eds. Recio-Blanco, de Laverny \& Brown

\bibitem[]{}Carollo, D., Beers, T. C., Placco, V. M. et al.: 2016
\textit{NatPh}, 12, 1170

\bibitem[]{} Cescutti, G.,  Chiappini, C.: 2014,
\textit{A\&A}, 565, A51

\bibitem[]{} Chiappini, C., Anders, F., Rodrigues, T. S. et al.: 2015,
\textit{A\&A}, 576, L12 
\bibitem[]{} Chiappini, C., Montalban, J., Steffen, M.: 2016,
\textit{AN}, 337, 773
\bibitem[]{} Chiappini, C., Matteucci, F., Gratton, R.: 1997,
\textit{ApJ}, 477, 765

\bibitem[]{} Debattista, V. P., Ness, M., Gonzalez, O. A. et al.: 2017,
\textit{MNRAS}, 469, 1587 

\bibitem[]{} Di Matteo, P., Haywood, M., Gomez, A. et al.: 2014
\textit{A\&A}, 567, A122

\bibitem[]{} Fragkoudi, F., Di Matteo, P., Haywood, M.  et al 2017,
\textit{A\&A}, sub (IAU S334, poster)
\bibitem[]{} Fuhrmann, K.: 2011,
\textit{MNRAS}, 414, 2893

\bibitem[]{} Garrison-Kimmel, S.,  Wetzel, A., Bullock, J. S.  et al 2017,
\textit{MNRAS}, sub, (arXiv)

\bibitem[]{} Hayden, M. R., Bovy, J., Holtzman, J. A. et al.: 2015,
\textit{ApJ}, 808, 132

\bibitem[]{} Haywood, M., Di Matteo, P., Lehnert, M. D. et al., 2013,
\textit{A\&A}, 560, A109

\bibitem[]{} Jofr\'e, P., Heiter, U., Worley, C. C. et al.: 2017,
\textit{A\&A}, 601, A38 

\bibitem[]{} Khoperskov, S., Haywood, M., Di Matteo, P., Lehnert, M., Combes, F.: 2017
\textit{A\&A}, sub (IAU S334, poster)

\bibitem[]{} Martig, M., Rix, H-W., Silva Aguirre, V. et al. 2015,
\textit{MNRAS}, 451, 2230 
\bibitem[]{} Martig, M., Minchev, I., Ness, M. et al. 2016,
\textit{ApJ}, 831, 139 

\bibitem[]{} Meynet, G., Ekstr\"om, S., Maeder, A.: 2006,
\textit{A\&A}, 447, 623
\bibitem[]{} Meynet, G., Hirschi, R., Ekstr\"om, S. et al.: 2010, 
\textit{A\&A}, 521, A30

\bibitem[]{} Miglio, A. et al.: 2017,
\textit{AN}, 338, 644
\bibitem[]{} Minchev, I., Martig, M., Streich, D. et al.: 2015,
\textit{ApJ}, 804, L9
\bibitem[]{} Minchev, I., Chiappini, C., Martig, M.: 2014,
\textit{A\&A}, 572, A92
\bibitem[]{} Minchev, I., Chiappini, C., Martig, M.: 2016,
\textit{AN}, 337, 944

\bibitem[]{} Oman, K. A., Marasco, A., Navarro, J. F. et al.: 2017, 
\textit{MNRAS}, sub (arXiv)

\bibitem[]{} Portail, M., Gerhard, O., Wegg, C., Ness, M.: 2017a,
\textit{MNRAS}, 465, 1621
\bibitem[]{} Portail, M., Wegg, C.,  Gerhard, O.,  Ness, M.: 2017b,
\textit{MNRAS}, 470, 1233

\bibitem[]{} Rahimi, A., Carrell, K., Kawata, D.: 2014,
\textit{RAA}, 14, 1406

\bibitem[]{} Recio-Blanco, A., Rojas-Arriagada, A., de Laverny, P. et al.: 2017,
\textit{A\&A}, 602, L14

\bibitem[]{} Robin, A. C., Bienaym\'e, O., Fernandez-Trincado, J. G., Reyl\'e, C.: 2017,
\textit{A\&A}, in press, (arXiv)

\bibitem[]{} Snaith, O. N., Haywood, M., Di Matteo, P., Lehnert, M., Combes, F., Katz, D., Gomez, A.: 2014, 
\textit{ApJ}, 781, L31
\bibitem[]{} Snaith, O. N., Haywood, M., Di Matteo, P., Lehnert, M., Combes, F., Katz, D., Gomez, A.: 2015,
\textit{A\&A}, 578, A87

\bibitem[]{} Soderblom, D.R.: 2010, ARA\&A 48, 581

\bibitem[]{} Wegg, C., Gerhard, O., Portail, M.: 2017,
\textit{ApJL}, in press (ArXiv)

\bibitem[]{} Xu, Y., Newberg, H. Jo, Carlin, J. L. et al. 2015,
\textit{ApJ}, 801, 105

\bibitem[]{} Zoccali, M., Vasquez, S., Gonzalez, O. A. et al. 2017,
\textit{A\&A}, 599, A12

\end{thebibliography}
\end{document}